\RequirePackage{fix-cm}
\documentclass[twocolumn,epjc3]{svjour3}  
\smartqed  % flush right qed marks, e.g. at end of proof
\RequirePackage{graphicx}
\newcommand{\be}{\begin{equation}}
\newcommand{\ee}{\end{equation}}

\journalname{Eur. Phys. J. C}
\begin{document}

\title{Lagrangian formulation of Omori's law and analogy with the 
cosmic Big Rip%\thanksref{t1}
}
%\subtitle{}


\author{Valerio Faraoni\thanksref{e1,addr1}
	%     \and
    	%Second Author\thanksref{e2,addr2,addr3} %etc.
}

\thankstext{e1}{e-mail: vfaraoni@ubishops.ca}


\institute{Department of Physics \& Astronomy, Bishop's University, 
2600 College Street, Sherbrooke, Qu\'ebec, Canada J1M~1Z7 \label{addr1}
%           Second address \label{addr2}
%           \emph{Present Address:} if needed\label{addr3}
}

\date{Received: date / Accepted: date}

\maketitle

\begin{abstract}

A recent model predicting Omori's law giving the number of aftershocks per 
unit time following an earthquake involves a differential equation 
analogous to the Friedmann equation of cosmology. The beforeshock phase is 
analogous to an accelerating universe approaching a Big Rip, the main 
shock to the Big Rip singularity, and the aftershock to a contracting 
universe. The analogy provides some physical intuition and Lagrangian and 
Hamiltonian formulations for Omori's law and its generalizations.

\end{abstract}

\section{Introduction}
\label{sec:1}

One of the first results obtained in modern seismology was Omori's law 
stating that, on 
average, following a strong earthquake the number of aftershocks per unit 
time $n(t)$ decays according to the empirical power 
law\footnote{The Omori law is used also to describe seismicity rates  
before and after 
eruptions \cite{Lemarchand,SchmidGrasso12}.} \cite{Omori} 
\be 
n(t)=\frac{k}{c+t}=\frac{k}{t-|c|} \,, \label{Omorisolution}
\ee 
where $k>0$ and $c<0$ are constants. There is a large body of literature 
on 
Omori's law (see \cite{Utsu,Guglielmireview} for reviews), but its 
physical interpretation is still 
mysterious, although it seems clear that somehow the 
source of the earthquakes should be traced to a rupture mechanism in 
the rocks composing the Earth's crust.
There is some belief that Omori's law is fundamental and not a mere  
data-fitting device and, in this optics, it makes sense to derive it from 
basic models.

The derivation proposed in Refs. 
\cite{Guglielmi0,Guglielmi,Guglielmireview} 
begins by noting that $n(t)$ satisfies the first order 
differential equation
\be
\dot{n}=-\sigma \, n^2 \,,\label{Omori} 
\ee
where $\sigma=k^{-1}$ and an overdot denotes differentiation with respect 
to time. The derivation uses an  analogy between the 
decaying number of aftershocks per unit time and the decreasing   density 
of ionospheric plasma due to the recombination of opposite charges 
\cite{Guglielmi0,Guglielmi,Guglielmireview}. If $n_{\pm}$ is the density of 
positive/negative charges and $n=n_{+}+n_{-}$, the recombination equation 
becomes $ \dot{n}=-\sigma n_{+} n_{-}$ and approximates to 
Eq.~(\ref{Omori}) for a globally neutral plasma in which $n_{+} \simeq 
n_{-}$.  Similarly, an earthquake occurs due the fast slip of rock along 
a fault plane in the Earth's crust, and there are two adjacent sides 
(denoted with $n_{+}$ and $n_{-}$) of a 
tectonic fault. Rupture releases the energy in an active fault and 
neutralizes the stresses on the parallel sides of it, reducing 
the number $n$ of active faults. The evolution of the number of faults 
then should obey \cite{Guglielmi0,Guglielmi,Guglielmireview}
\be
\frac{dn}{dt} = -\sigma n_{+} n_{-} \simeq -\sigma n^2 
\ee
where $\sigma$ is a deactivation coefficient and 
$n_{+}=n_{-}$ has been used. The fact that a pair of adiacent fault sides 
is 
involved rules out different powers in the Omori law\footnote{In 
principle, however, the deactivation coefficient $\sigma$ could depend 
on time, introducing nonstationarity and deviations from a strict Omori 
law \cite{Guglielmi,Guglielmireview}.} \cite{Guglielmi,Guglielmireview}. 

The beforeshock phase, during which secondary shocks increase their 
frequency until the main shock, can be described by a version of Omori's 
law $\dot{n}= \sigma \, n^2$, although the phenomenological descriptions  
and data fitting are different. Here we point out that there are many 
similarities between the differential equation satisfied by Omori's law 
and the Friedmann equation of spatially homogeneous and isotropic 
Friedmann-Lema\^itre-Robertson-Walker (FLRW) cosmology 
\cite{Landau,Carroll,Wald,Peebles,Liddle,KT}. The analogy holds 
in the case of a universe with a phantom fluid as the matter source and 
with a Big Rip singularity occurring at a finite time. The Big Rip 
separates the 
``before'' and ``after'' universes and is analogous to the main earthquake 
shock. This analogy is intriguing and may provide some physical intuition 
about  
variability of the deactivation coefficient $\sigma$ versus variability of 
the power in Omori's law. What is more, the 
analogy reveals previously unknown Lagrangian and Hamiltonian formulations 
of the physical system described by the Omori law~(\ref{Omori}) and its 
generalizations. 

In the next section we discuss the Lagrangian and Hamiltonian associated 
with the Omori law~(\ref{Omori}). In Sec.~\ref{sec:3} we recall the basics 
of FLRW cosmology and we present the analogy with a Big Rip in a  
spatially flat universe, while Sec.~\ref{sec:4} contains the conclusions.

\section{Lagrangian formulation of Omori's law and a mechanical analogy}
\label{sec:2}

It is not obvious that Omori's law can be described using the Lagrangian 
or Hamiltonian formalisms. A Lagrangian leading to Omori's law is
\be
L\left( n, \dot{n} \right)= n\dot{n}^2 + \sigma^2 \, n^5 \,. 
\label{Lagrangian}
\ee
In fact, the Euler-Lagrange equation
\be
\frac{d}{dt} \left( \frac{\partial L}{\partial \dot{n}} 
\right)-\frac{\partial L}{\partial n}=0
\ee
yields
\be
2n\ddot{n} +\dot{n}^2 -5\sigma^2 n^4 =0 \,. \label{eq:cio}
\ee
Now, the Omori law~(\ref{Omori}) is  a first integral of 
Eq.~(\ref{eq:cio}). In fact, by differentiating~(\ref{Omori}) one obtains
\be
\ddot{n}=-2\sigma \, n \dot{n}=2\sigma^2 \, n^3 \,,
\ee
using which one verifies that $ 2n\ddot{n}+\dot{n}^2-5\sigma^2 
n^4 =0$. 

The corresponding Hamiltonian is
\be
{\cal H} = \pi_n \dot{n}-L = n \left( \dot{n}^2 -\sigma^2 n^4 \right) 
\,,\label{Hamiltonian}
\ee
where $\pi_n \equiv \partial L/\partial \dot{n} = 2n\dot{n}$ is the 
momentum 
canonically conjugated to the variable $n$. One notes that $\partial 
{\cal H}/\partial t$ vanishes and the Hamiltonian is conserved, ${\cal 
H}=$~const. Furthermore, using the Omori law~(\ref{Omori}) in 
Eq.~(\ref{Hamiltonian}) gives
\be
{\cal H}=0 \,,\label{zeroHamiltonian}
\ee
{\em i.e.}, the point-particle system associated with the Omori 
Lagrangian and Hamiltonian has conserved total energy equal to zero. 

One can write 
\be
\frac{\cal H}{2} = \mu \left( \frac{\dot{n}^2}{2} - \frac{\sigma^2}{2} \, 
n^4 \right)
\ee
where, for $n \geq 0$,  $\mu(n)=n $ is a position-dependent mass, with 
kinetic 
energy $\mu \, \dot{n}^2/2$, potential energy 
$ V(n)=-\mu \sigma^2 n^4/2$, and zero total mechanical energy.   Since 
$\dot{n}<0$, the particle will move to the 
left of the $n$-axis, tending toward $n=0$ ({\em i.e.}, the 
seismic activity is more intense at the initial point $n_{(0)}>0$ and 
stops at $n=0$).

The $\left( n , \dot{n} \right)$ phase plane associated with Omori's law 
has a very simple structure. Equation~(\ref{Omori}) or, equivalently, 
Eq.~(\ref{zeroHamiltonian}) is an energy constraint that reduces the 
orbits of the solutions to move on the parabolas $\dot{n}(n) = \mp \sigma 
n^2$, with the upper sign corresponding to the aftershock phase and the 
lower one to the beforeshock phase. The two parabolas correspond to the 
orbits of two different dynamical systems and are considered here as 
living in the same phase plane only for convenience: the fact that they 
touch each other at the origin $(0,0)$ has no meaning since these are 
disconnected curves.

The aftershock phase corresponds to the lower quadrant $n \geq 0, 
\dot{n}\leq 0$, in which the point representing the state of the system 
moves along the downward-facing parabola towards the origin, which is an 
attractor. In this regime, secondary shocks decay in a finite time $|c|$.

The beforeshock phase corresponds to the upper quadrant $n\geq 0, 
\dot{n}\geq 0$, in which the point representing the dynamical system 
moves away from the origin and upward toward infinite $n$ and $\dot{n}$, 
reaching infinity in a finite time. The main shock corresponds to infinity 
in this plane, to the pole $t=|c|$ in the solution 
\be
n(t)=\frac{k}{ \left|t -|c|\right|} \,,
\ee
and to a discontinuity in the dynamics.

\section{Analogy with a cosmic Big Rip singularity}
\label{sec:3}

One can square Eq.~(\ref{Omori}) and rewrite it as
\be
\left(\frac{\dot{n}}{n}\right)^2 = \sigma^2 n^2 \label{Omorisquared}
\ee
which is analogous to the Friedmann equation of cosmology if one exchanges 
$n(t)$ with the cosmic scale factor. In order to develop the analogy, let 
us  recall the  basics of FLRW cosmology 
\cite{Carroll,Wald,Peebles,Liddle,KT}.

In general relativity \cite{Landau,Carroll,Wald}, a  spatially 
homogeneous and 
isotropic universe can only have one of three possible geometries, which 
are described by the four-dimensional FLRW line 
element given, in comoving polar coordinates $\left(t, r, \theta, \varphi 
\right)$, by
\begin{eqnarray}
ds^2 &=& -dt^2 +a^2(t) \left[ \frac{dr^2}{1-Kr^2} +r^2 \left( d\theta^2 + 
\sin^2 \theta \, d\varphi^2 \right)\right] \,.\nonumber\\
&& \label{eq:10}
\end{eqnarray}
The function $a(t)$ (``scale factor'') quantifies how two points at 
fixed comoving distance $r_0$ ({\em e.g.}, two  average 
galaxies without proper motions) move away from each other as the universe 
expands. Their  physical separation at time $t$ is $ 
l(t)=a(t)r_0$ and it  increases in an expanding universe described by 
increasing $a(t)$. Therefore, the scale factor $a(t)$ illustrates the 
expansion history of the universe.

The constant $K$ in 
Eq.~(\ref{eq:10}) is normalized to $K= 1, 0, -1$ corresponding, 
respectively, to a closed universe (closed 
three-dimensional  spatial sections $t=$~const.), Euclidean spatial 
sections, or hyperbolic 
3-spaces \cite{Landau,Carroll,Wald,Peebles,Liddle,KT}, which 
includes 
all the possible FLRW geometries. The cosmic dynamics is described  
by $a(t)$ \cite{Landau,Carroll,Wald,Peebles,Liddle,KT}.

In relativistic cosmology the matter content of the universe, 
which is the source of the spacetime curvature, is usually modelled by a 
perfect fluid with energy 
density $\rho(t)$ and isotropic pressure $P(t)$. These quantities are 
related by some equation of state, usually (but not necessarily) of the 
form $P=w\rho$ with 
$w=$~const.

The functions $a(t), \rho(t)$, and $P(t)$ obey the 
Einstein-Friedmann equations
\begin{eqnarray}
&&H^2 \equiv \left( \frac{\dot{a}}{a}\right)^2 =\frac{8\pi G}{3} \, \rho 
-\frac{K}{a^2} \,, \label{eq:11}\\
&&\nonumber\\
&&\frac{\ddot{a}}{a}= -\, \frac{4\pi G}{3} \left( \rho +3P \right) \,, 
\label{eq:12} \\
&&\nonumber\\
&& \dot{\rho}+3H\left(P+\rho \right)=0 \,,\label{eq:13}
\end{eqnarray}
where $G$ is Newton's constant, units in which the speed of 
light is unity are used, differentiation with respect to the comoving time  
$t$ is denoted by an overdot, and  
$H(t)\equiv  \dot{a}/a$ is the Hubble function 
\cite{Carroll,Wald,Peebles,Liddle,KT}. 
There are only two independent equations in the 
set~(\ref{eq:11})-(\ref{eq:13}) 
since any one of them can be derived from the other two.  Without losing  
generality, we choose the Friedmann equation~(\ref{eq:11}) and the energy 
conservation equation~(\ref{eq:13}) as independent, then the acceleration 
equation~(\ref{eq:12}) follows from them. 

Equation~(\ref{eq:11}) with $K=0$ is formally the same as 
the squared Omori differential equation~(\ref{Omorisquared}) under the 
exchange 
$ n(t) \longrightarrow a(t) $ provided that the analogous  
universe is sourced by a suitable 
cosmological fluid.  Equations~(\ref{eq:11}) 
and~(\ref{Omorisquared}) considered jointly imply that it   must be
\be
\rho(t)=  \rho_0 a^2(t)  
\,,\label{eq:14} 
\ee 
where $\rho_0 $ is a positive integration constant determined by the 
initial conditions and such that 
\be
\sigma^2 = \frac{8\pi G \rho_0}{3} \,.\label{questa}
\ee 
In FLRW cosmology, where the cosmic fluid satisfies 
the barotropic equation 
of  state $ P=w\rho $,  $w=$~const., Eq.~(\ref{eq:13})  integrates  
immediately to
\be
\rho(a) = \frac{ \rho_0}{ a^{3(w+1)} }  \label{eq:16} \,.
\ee
The corresponding solution of the Friedmann 
equation is 
\be
a(t)=\frac{a_0}{ \left| t-t_0\right|^{3|w+1| } } \,.
\ee
The comparison of Eqs.~(\ref{eq:14}) and~(\ref{eq:16}) shows that the 
analogy between earthquakes and cosmology is valid if the universe is 
filled with a perfect  fluid with $P=w\rho$ and equation of state 
parameter $ w= -5/3$.

The aftershock regime corresponds to a contracting universe with 
decreasing $a(t)$ and $\dot{n}<0$, while the beforeshock phase corresponds 
to an expanding analogous universe and $\dot{n}>0$.

It is well known \cite{Carroll,Wald} that the Friedmann equation is a 
first order constraint and not a truly dynamical (second order) equation 
of motion. This constraint (``Hamiltonian constraint'') corresponds to the 
vanishing of the Hamiltonian of general relativity \cite{Carroll,Wald,KT},  
and this is exactly the role played by the law~(\ref{Omori}), as seen 
in Eq.~(\ref{Hamiltonian}). The facts that the Friedmann equation looks 
like an energy conservation equation for one-dimensional motion and that 
it can describe a variety of different universes makes it suitable for 
several analogies between the cosmos and unrelated physical systems, 
including Bose-Einstein condensates \cite{BEC}, glacial valleys 
\cite{profiles}, capillary fluids \cite{capillary}, equilibrium beach 
profiles \cite{beach}, and freezing bodies of water \cite{freezinglakes}.

In the aftershock phase with $\dot{n}<0$, the analogous Friedmann equation 
describes a spatially flat ($K=0$) contracting universe fueled by a 
perfect fluid with energy density $\rho =\rho_0 a^2$ and equation of state 
parameter $w= -5/3$. This ``phantom fluid'' violates 
all the energy conditions expected to hold for physically reasonable  
matter \cite{Carroll,Wald,Peebles,Liddle}. Nevertheless, phantom matter is 
the subject of a large body of literature in cosmology because it can 
potentially explain a superaccelerating ({\em i.e.}, $\dot{H}>0$) universe 
often preferred by cosmological observations.

A peculiar feature of a phantom fluid is that it causes a universe filled 
with it to expand so fast that it explodes at a finite time in a Big Rip 
singularity \cite{BigRip}. Contrary to the better known Big Bang or Big 
Crunch 
singularities where the scale factor vanishes, in a Big Rip $a(t)$ 
diverges. Scalar curvature invariants, as well as the energy density 
$\rho$ and the pressure $P$ also diverge, making the Big Rip a genuine 
spacetime singularity \cite{BigRip}.

In our analogy, $t = |c|$ corresponds to the main earthquake shock and is 
 analogous to the Big Rip singularity, while the aftershock 
phase $\dot{n}<0$ corresponds to the less studied branch of a universe 
{\em 
contracting} from a Big Rip. The expanding and contracting 
branches on either side of the Big Rip are disconnected because  a 
spacetime manifolds stops at a curvature 
singularity (in this case, the Big Rip), which is not part of spacetime 
itself. The expanding branch of the phantom universe has an analog in the 
Omori law with sign changed, $\dot{n}=\sigma n^2$, which can be used 
to model the beforeshock phase of an earthquake, 
during which smaller 
shocks become more and more frequent and lead to the main shock 
\cite{Lemarchand,SchmidGrasso12}. 
The main earthquake separating beforeshock and aftershock regimes is 
analogous 
to the Big Rip singularity.

\section{Discussion and Conclusions}
\label{sec:4}

We have developed an analogy between Omori's law for the aftershocks 
following a main earthquake event and a spatially flat universe in FLRW 
cosmology, which is sourced by a phantom fluid and contracting. It is 
natural to extend this analogy to include a beforeshock phase 
corresponding to an expanding universe sourced by the same (or 
another) phantom fluid. The Big 
Rip singularity separating the expanding and contracting universes 
is analogous to the spacetime singularity.

Formally, the catastrophic nature of the solution of Eq.~(\ref{Omori}) 
and of $\dot{a} \approx a^2$ is due the fact that the exponent~2 in the 
right hand side prevents the existence of a maximal solution defined of an 
infinite half-interval \cite{BrauerNoel}.  The analogy has  some value for 
physical 
intuition. Indeed, the Lagrangian~(\ref{Lagrangian})  for Omori's law 
is derived using as an example  the effective point-like Lagrangian for a 
FLRW 
universe sourced by a perfect fluid, which is ({\em 
e.g.}, \cite{cosmoLagrangian,mybook})
\be
L\left( a, \dot{a}, P \right)= 3a\dot{a}^2 -a^3 P  \,.
\ee

Another consideration is in order. Aftershocks are often modelled with the 
generalized Omori (or Omori-Utsu) law \cite{Hirano24,Jeffreys38}
\be 
n(t) = \frac{k}{ \left(  t-|c|\right)^p}\,, 
\ee 
where the exponent $p$ varies according to the 
location and the specific earthquake in a rather wide range \cite{Utsu61}.  
In this case the analog of Eq.~(\ref{Omori}) is 
\be 
\dot{n} 
=-\frac{p}{k^p} \, n^{\frac{p+1}{p} } \equiv -\sigma_{(p)} 
n^{\frac{p+1}{p}} \,. 
\ee
One can generalize the previous reasoning for $p=1$: the  
Lagrangian is now
\be
L_{(p)} \left(n, \dot{n}\right) = n \dot{n}^2 +\sigma_{(p)}^2 \, 
n^{\frac{3p+2}{p}} \,,
\ee
the second order equation of motion is 
\be
2n\ddot{n}+\dot{n}^2 -\frac{(3p+2)}{p} \, \sigma_{(p)}^2 \, n^{ 
\frac{2(p+1)}{p}} 
\,,
\ee
while the  Hamiltonian is 
\be
{\cal H}_{(p)}= n \left( \dot{n}^2 - \sigma_{(p)}^2 \, 
n^{\frac{2(p+1)}{p}} \right) \,;
\ee
it is conserved, and its value is again ${\cal H}_{(p)}=0$. The analogy 
with cosmology is still valid and, for the range of values of $p>0$  
encountered in the literature, the cosmic fluid is again a phantom 
fluid with equation of state parameter
\be
w_{(p)} = - \frac{(3p+2)}{3p}  
\ee
causing again a Big Rip (which always occurs for equation of state 
parameters $w<-1$ \cite{BigRip}).

In principle, a deviation of the 
exponent $p$ from unity ruins the simple derivation of 
Refs.~\cite{Guglielmi0,Guglielmi,Guglielmireview}. These authors  
attribute deviations from 
the simple Omori law (\ref{Omorisolution}) to a time dependence of the 
coefficient $\sigma$ instead. In the cosmological analogy, a varying 
$\sigma$ corresponds to a time-varying gravitational constant $G$ (cf. 
Eq.~(\ref{questa})), which 
is impossible in general relativity. Such a variation is an essential part 
of scalar-tensor cosmology, but this possibility necessarily implies the 
presence of additional terms in the Friedmann and acceleration 
equations~(\ref{eq:11}) and (\ref{eq:12}) 
\cite{BD,mybook,FujiiMaeda,Salvbook}. The lesson from cosmology would be 
that the variation of $\sigma$ involves extra energy terms associated 
with $\dot{\sigma} \neq 0$ in an energy 
balance involving the variation of $n$.  
It is more natural, and common in the cosmological literature, to allow 
for a different equation of state parameter or, perhaps, for 
time-dependent equation of state of the cosmic fluid $P(t)=w(t) 
\rho(t)$. This would still be a perfect fluid and can be realized, 
for example, 
by a scalar field with a dynamical equation of state, as in early universe 
inflation \cite{inflation,KT,Liddle} and the late time, dark 
energy-dominated, era 
\cite{AmendolaTsujikawa}. Both procedures would imply the introduction of  
another element in  the fundamental derivation of the Omori law of 
Refs.~\cite{Guglielmi0,Guglielmi,Guglielmireview},  perhaps a distribution 
of intersecting faults with more 
than two adiacent sides involved. Here we do not speculate further on this 
new element. In any case, the search for fundamental and universal laws as 
opposed to mere data-fitting lies at the core of science. Lagrangian 
and Hamiltonian formulations and analogies can perhaps help in the search 
for these laws.

\begin{acknowledgements}

This work is supported, in part, by Bishop's University and by the Natural 
Sciences \& Engineering Research Council of Canada (Grant No. 2016-03803).

\end{acknowledgements}


\begin{thebibliography}{}

\bibitem{Omori} F. J. Omori, ``On the Aftershocks of Earthquakes'', {\em 
J. College Sci. Imperial Univ. Tokyo} {\bf 7}, 111 (1894) 

\bibitem{Lemarchand} N. Lemarchand and J.-R. Grasso, ``Interactions 
between earthquakes and volcano activity'', {\em Geophys. Res. Lett.} {\bf 
34}, L24303 (2007) 
	%doi:10.1029/ 2007GL031438.

\bibitem{SchmidGrasso12} A. Schmid and J.-R. Grasso, ``Omori law for 
eruption foreshocks and aftershocks'', {\em J. Geophys. Res. } {\bf 
117}, B07302 (2012) 

\bibitem{Utsu} T. Utsu, Y. Ogata, and R. S. Matsu'ura, ``The centenary of 
the Omori formula for a decay law of aftershock activity,  {\em  J. Phys. 
Earth} {\bf 43}, 1 (1995) 
	%1-33

\bibitem{Guglielmireview} A. V. Guglielmi, ``Omori's law: a note on the 
history of geophysics'', {\em Physics Uspekhi} {\bf 60}, 319 (2017) 
	%-324

\bibitem{Guglielmi0} A. V. Guglielmi, ``Interpretation of the Omori law'', 
{\em Izv., Phys. Solid Earth} {\bf 52}, 785 (2016) %–786
	%https://doi.org/10.1134/S1069351316050165

\bibitem{Guglielmi} A. V. Guglielmi and A. D. Zavyalov, ``The 150th 
Anniversary of Fusakichi Omori'', arXiv:1803.08555.

\bibitem{Landau} L. D. Landau and E. M. Lifschitz, {\em The Classical 
Theory of Fields}. Pergamon, Oxford (1989)

\bibitem{Carroll} S. M. Carroll, {\em Spacetime and Geometry: An 
Introduction to General Relativity}. Addison Wesley, San Francisco (2004).

\bibitem{Wald} R. M. Wald, {\it General Relativity}. Chicago University 
Press, Chicago (1984)

\bibitem{Peebles} P. J. E. Peebles, {\em Principles of Physical 
Cosmology}. Princeton University Press, Princeton (1993)

\bibitem{Liddle} A. Liddle, {\em An Introduction to Modern 
Cosmology}. Wiley, Chichester (2003)

\bibitem{KT} E. W. Kolb and M. S. Turner, {\em The Early 
Universe}. Addison-Wesley, Redwood City, CA (1990)

\bibitem{BEC} P. O. Fedichev and U. R. Fischer, ``Gibbons-Hawking Effect 
in the Sonic de Sitter Space-Time of an Expanding Bose-Einstein-Condensed 
Gas'', {\em Phys. Rev. Lett.} {\bf 91}, 240407 (2003) ; {\em Erratum, 
Phys. Rev. Lett.} {\bf 92}, 049901(E) (2004); C. Barcelo, S. Liberati, and 
M. Visser, ``Analog models for FRW cosmologies'', {\em Int. J. Mod. Phys. 
D} {\bf 12}, 1641 (2003); P. O. Fedichev and U. R. Fischer, 
````Cosmological'' quasiparticle production in harmonically trapped 
superfluid gases'', {\em Phys. Rev. A} {\bf 69}, 033602 (2004); U. R. 
Fischer and R. Sch\"utzhold, ``Quantum simulation of cosmic inflation in 
two-component Bose-Einstein condensates'', {\em Phys. Rev. A} {\bf 70}, 
063615 (2004); S.-Y. Ch\"a and U. R. Fischer, ``Probing the Scale 
Invariance of the Inflationary Power Spectrum in Expanding 
Quasi-Two-Dimensional Dipolar Condensates'', {\em Phys. Rev. Lett.} {\bf 
118}, 130404 (2017); {\em Erratum Phys. Rev. Lett.} {\bf 118}, 179901(E) 
(2017); S. Eckel, A. Kumar, T. Jacobson, I. B. Spielman, and G. K. 
Campbell, ``A Rapidly Expanding Bose-Einstein Condensate: An Expanding 
Universe in the Lab'', {\em Phys. Rev. X} {\bf 8}, 021021 (2018).

\bibitem{profiles} S. Chen, G. W. Gibbons, and Y. Yang, ``Explicit 
integration of Friedmann’s equation with nonlinear equations of state'', 
{\em J. Cosmol. Astropart. Phys.} {\bf 05}, 020 (2015); S. Chen, G. W. 
Gibbons, 
and Y. Yang, ``Friedmann-Lemaitre cosmologies via roulettes and other 
analytic methods'', {\em J. Cosmol. Astropart. Phys.} {\bf 10},  
056 (2015); 
V. Faraoni and A. M. Cardini, ``Analogues of glacial valley profiles in 
particle mechanics and in cosmology'', {\em FACETS} {\bf 2}, 286 (2017)
	%–300 DOI: 10.1139/facets-2016-0045

\bibitem{capillary} D. Bini and S. Succi, ``Analogy between capillary 
motion and Friedmann-Robertson-Walker cosmology'', {\em Europhys. Lett.} 
{\bf 82}, 34003 (2008) 
	%doi: 10.1209/0295-5075/82/34003

\bibitem{beach} V. Faraoni, ``Analogy between equilibrium beach 
profiles and closed universes'', {\em Phys. Rev. Research} {\bf 1},  
033002 (2019)

\bibitem{freezinglakes} V. Faraoni, ``Analogy between freezing lakes and 
the cosmic radiation era'', {\em Phys. Rev. Research} {\bf 2}, 
013187 (2020) 

\bibitem{BigRip} R. R. Caldwell, ``A phantom menace? Cosmological 
consequences of a dark energy component with super-negative equation of 
state'', {\em Phys. Lett. B} {\bf 545}, 23 (2002); 
R. R. Caldwell, M. Kamionkowski, and N. N. Weinberg, ``Phantom energy 
and cosmic doomsday'', {\em Phys. Rev. Lett.} {\bf 91},  071301 (2003)

\bibitem{cosmoLagrangian} S. Capozziello and R. de Ritis,``Relation 
between the potential and nonminimal coupling in inflationary cosmology'', 
{\em Phys. Lett. A} {\bf 177}, 1 (1993); ``N\"other's symmetries and 
exact 
solutions in flat non-minimally coupled cosmological models'', {\em Class. 
Quantum Grav.} {\bf 11},  107 (1994); M. Demianski, R. de Ritis, G. 
Platania, C. Rubano, P. Scudellaro, and C. Stornaiolo, ```Scalar field, 
nonminimal coupling, and cosmology'', {\em Phys. Rev. D} {\bf 44},  
3136 (1991); A. K. Sanyal and B. Modak, ``Is Noether Symmetric Approach 
Consistent With Dynamical Equation In Non-minimal Scalar-Tensor 
Theories?'', {\em Class. Quantum Grav.} {\bf 18}, 3767 (2001) 

\bibitem{mybook} V. Faraoni, {\em Cosmology in Scalar-Tensor 
Gravity}. Kluwer Academic, Dordrecht (2004)

\bibitem{BrauerNoel} F. Brauer and J.A. Noel, {\em Introduction to 
Differential Equations  With Applications}. Harper \& Row, New York 
(1986)

\bibitem{Hirano24} S. Hirano, ``Investigation of aftershocks of the great 
Kanto earthquake at Kumagawa'', {\em Kishosushi, Ser.} 2, {\bf 2},  
77 (1924) (in Japanese)

\bibitem{Jeffreys38} H. Jeffreys, ``Aftershocks and periodicity in 
earthquakes'', {\em Gerlands Beitr. Geophys.} {\bf 86}, 111 (1938)

\bibitem{Utsu61} T. Utsu, ``Statistical study on the occurrence of 
aftershocks'', {\em Geophys. Mag.} {\bf 30},  521 (1961)

\bibitem{BD} C. Brans and R.H. Dicke, ``Mach's principle and a 
relativistic theory of gravitation'', {\em Phys. Rev.} {\bf 124}, 
925 (1961) 

\bibitem{FujiiMaeda} Y. Fujii and K. Maeda, {\em The Scalar-Tensor Theory 
of Gravitation}. Cambridge University Press, Cambridge (2003)

\bibitem{Salvbook} S. Capozziello and V. Faraoni, {\em Beyond Einstein 
Gravity: A Survey of Gravitational Theories for Cosmology and 
Astrophysics}. Springer, New York (2010)

\bibitem{inflation} A. A. Starobinsky, ``A new type of isotropic 
cosmological models without singularity'', {\em Phys. Lett. B} {\bf 91}, 
99 (1980); A. H. Guth, ``Inflationary universe: A possible solution to 
the horizon and flatness problems'', {\em Phys. Rev. D} {\bf 23},  
347 (1981); A. Linde, {\em Particle Physics and Inflationary 
Cosmology}. Harwood 
Academic, Chur, Switzerland (1990); V. Mukhanov, {\em Physical Foundations 
of Cosmology} (Cambridge University Press, Cambridge (2005)

\bibitem{AmendolaTsujikawa} L. Amendola and S. Tsujikawa, {\em Dark 
Energy, Theory and Observations}. Cambridge University Press, Cambridge 
(2010)

\end{thebibliography}


\end{document}